\documentstyle[aps,preprint,epsfig]{revtex}

\begin{document}
\tighten

\preprint{SNUTP-99-10}
\title{Gravitino as a Sterile Neutrino}
\author{Taekoon Lee}
\address{Center for Theoretical Physics, Seoul National University\\
Seoul 151-742, Korea}
\maketitle
\begin{abstract}
A scheme is described  in which the light gravitino in low energy SUSY
breaking models mixes with neutrinos.
The mixing between gravitino and neutrinos arises through the standard model
symmetry breaking and an R-parity and lepton number violating bilinear term 
in the superpotential. It is shown that mixings compatible with the neutrino
experiments can be obtained within the cosmological bound on the bilinear term
set by the baryon asymmetry of the universe.
\end{abstract}

The most promising explanation for the solar and atmospheric
neutrino flux deficit is the neutrino oscillation that arises when
there are mixings among a group of fermions that includes the three
families of neutrinos and possibly sterile fermions  neutral under
the standard model symmetry group. Although the scheme in which
mixings occur only within the three families of
neutrinos is most economic, neutrino mixings with sterile fermions
are an interesting possibility \cite{STERILE0,STERILE1}.
Moreover, the mixings with a sterile neutrino becomes {\it necessary} when one
attempts to explain the combined data of the solar, atmospheric and LSND 
neutrino experiments which suggest existence of three different oscillation
lengths. Even if not all of these neutrino experiments turn out
correct we can still think of  neutrino oscillation with sterile fermions.

The sterile fermions that mix with neutrinos must be light $\leq O(1
\,\mbox{eV})$ to 
be relevant for neutrino oscillation. Some of the candidates
for the sterile fermions
considered in the literature include axino and  
modulinos \cite{Chun0,Chun,Smirnov}.
Being the superpartners of light scalars (axion and  moduli,
respectively) these fermions get small masses. In the low energy 
supersymmetry (SUSY)
breaking models such as the  gauge mediated SUSY breaking \cite{Dine}
or the no-scale supergravity \cite{Nano},
the gravitino can also be light \cite{Yanagida,Nanopoulos}, and so could 
play the role of a 
sterile neutrino. In this letter we consider mixings of light
gravitino with neutrinos.

Gravitino is the spin $\frac{3}{2}$  superpartner of graviton and
becomes massive when it absorbs through the super-Higgs mechanism
the Goldstino from the 
spontaneous global SUSY breaking. 
The gravitino interaction with matter is dominated by its spin $\frac{1}{2}$ 
longitudinal component, which is essentially the absorbed Goldstino, and the
interaction strength depends primarily on the 
SUSY breaking scale whereas the transverse components interaction with
matter is suppressed by Plank mass. Thus the mixings between gravitino and 
neutrinos can arise through the Goldstino-neutrino mixings.
 
The easiest way to describe gravitino-neutrino mixing in supergravity
is using the equivalence theorem \cite{Equiv}
to replace the gravitino with Goldstino
and work with the corresponding globally supersymmetric lagrangian.
The equivalence theorem is applicable in our case
since the neutrino energy  is much larger than the gravitino mass in
consideration.

Our scheme for the Goldstino-neutrino mixing is as follows.
Mixings between Goldstino and neutrinos can occur only when the electroweak
symmetry is broken since Goldstino is neutral under the symmetry. When
the electroweak symmetry is broken the Goldstino necessarily mixes with the
Higgsinos. To convert this mixing to Goldstino-neutrino mixing we introduce
a small R-parity and lepton number violating bilinear term in the superpotential
through which a neutrino-Higgsino mixing arises, which in turn mediates the
Goldstino-neutrino mixing.

In our scheme the bilinear terms in the superpotential are given by
\begin{equation}
W= \epsilon_{i} L_{i} H_{2} + \mu H_{1} H_{2}
\end{equation}
where $i=e,\mu,\tau$ and $L_{i}, H$ denote the lepton and Higgs
chiral fields, respectively.
The first term violates R-parity and lepton number conservation.
There is a strong constraint on the magnitude of $\epsilon_{i}$ from
the baryon asymmetry of the universe. When $\epsilon_{i}$ are sizable,
the lepton number violation combined with the $B+L$ violation through
the sphaleron transitions washes out any relic $B-L$ inherited
before the weak symmetry
breaking and results in $B=L=0$ universe. The constraint
from this consideration is given by \cite{Fukugita-Yanagida,Campbell,Ma}
\begin{equation}
\epsilon_{i} \leq 10^{-6} \,\,\mbox{GeV}.
\label{eq2}
\end{equation}
We shall see that a gravitino-neutrino mixing can arise within this
bound.

When the electroweak symmetry is broken, the neutral Higgsinos mix with
the Goldstino with a mixing angle of $O( F_{2}/F)$ where $F_{2}=\mu H_{1}$
and $F$ is the Goldstino decay constant.
Thus the Higgsino $\widetilde{H}_{2}$ can be written
in mass eigenstates as
\begin{equation}
\widetilde{H}_{2}\sim \frac{F_{2}}{F} \chi + \cdots
\end{equation}
where $\chi$ denotes  the Goldstino.
The Goldstino-Higgsino  mixing and the neutrino-Higgsino mixing from
the R-parity violating interaction gives the Goldstino-neutrino mixing
\begin{eqnarray}
{\cal L}_{\nu_{i}\chi} &\sim& m_{\nu_{i}\chi} \nu_{i}\chi + c.c.
\label{e5}
\end{eqnarray}
with 
\begin{equation}
 m_{\nu_{i}\chi}= \frac{\epsilon_{i}\mu H_{1}}{F}=  
 \frac{\epsilon_{i}\mu v \cos\beta}{F}\sim 2\left(\frac{\epsilon_{i}}{10^{-7}
 \mbox{GeV}}\right)\left(\frac{\mu}{300\,\mbox{GeV}}\right)
 \left(\frac{2\,\mbox{TeV}}{\sqrt{F}}\right)^{2}\cos\beta\,\, 
 \mbox{(eV)}
\end{equation}
and
\begin{equation}
\tan\beta=\frac{H_{2}}{H_{1}}.
\end{equation}

When the gravity is coupled, the Goldstino becomes the longitudinal 
component of the gravitino and obtains a mass 
\begin{equation}
m_{\chi}=
\frac{F}{\sqrt{3}M_{P}}\sim 10^{-3}
\left(\frac {\sqrt{F}}{2\,\mbox{TeV}}\right)^{2} \,\,\mbox{(eV)}
\end{equation}
where $M_{P}$ is the Plank mass.

Let us now see that the mixing given in (\ref{e5})
can be compatible with the neutrino
experiments. The neutrino masses may arise
through the seesaw mechanism or others,
but here we do not make any speculation  on the
pattern of the neutrino mass matrix. Instead we  treat neutrino masses as
parameters, and show that  for a certain
range of  neutrino masses the above Goldstino-neutrino mixing can account for
some of the missing neutrinos observed in the neutrino experiments.
In the following we take $\mu=300\,\mbox{GeV}$ and $\cos\beta=1/\sqrt{2}$
as a  reference.

The just-so solution for the solar neutrino problem with large
mixing angle and $\Delta m^{2}\sim 10^{-10} \,\mbox{eV}^{2}$ would be
achieved when $m_{\chi}, m_{\nu_{e}} \leq m_{\nu_{e}\chi} \approx 10^{-5}$ eV
which can be satisfied if $\sqrt{F} \sim 200$ GeV and $
\epsilon_{e} \sim 10^{-14}$ GeV are taken. Note that the SUSY breaking scale
$\sqrt{F}$ is  small but
still  above the low bound from the primordial nucleosynthesis
\cite{Tony}, and $\epsilon_{e}$ is well below the cosmological bound
(\ref{eq2}). The MSW small mixing solution for 
the solar neutrino puzzle requiring
a mixing angle $\theta \approx 4 \times 10^{-2}$ and 
$ \Delta m^{2}\approx 4\times 10^{-6} \mbox{eV}^{2}$  would be
realized provided $m _{\nu_{e}} < m_{\chi} \approx 2\times 10^{-3}$ eV,
$m _{\nu_{e}\chi} \approx 10^{-4}$ eV. This can be satisfied if we take
$ \sqrt{F} \sim 2$ TeV and $\epsilon_{e}\sim 10^{-11}$ GeV which is also
well below the bound (\ref{eq2}).
Now the atmospheric neutrino data from Super-Kamiokande collaboration
suggest a neutrino oscillation of almost maximal mixing and
$\Delta m^{2}\approx 2 \times 10^{-3} \mbox{eV}^{2}$. 
This would be realized  if $m_{\nu_{\mu}\chi} \gg m_{\chi}$ and
$ m_{\chi} \geq m_{\nu_{\mu}}$ are satisfied. 
Then the  maximal mixing arises from  the
pseudo Dirac nature of the neutrino 
mass matrix, and the mass splitting is given by
\begin{eqnarray}
\Delta m^{2}\sim 2 m_{\nu_{\mu}\chi}m_{\chi} 
\approx 3\times 10^{-3} \left(\frac{\epsilon_{\mu}}{10^{-7}\mbox{GeV}}\right)
\,\, \mbox{(eV}^{2})
\end{eqnarray}
which would give the required value if
$\epsilon_{\mu}\sim 10^{-7}$ GeV is taken. Note  again that $\epsilon_{\mu}$
satisfies the cosmological bound.
It is remarkable that
the mass splitting is independent of the SUSY breaking scale
except the condition $\sqrt{F} \leq 20$ TeV obtained by
assuming $ m_{\chi}/m_{\nu_{\mu}\chi} \leq 10^{-1}$.

The combined data of the  solar, atmospheric and LSND neutrino experiments
suggest existence of three different $\Delta m^{2}$ which is possible only
when there are more than four neutrinos. In this case the mixing of a
sterile neutrino with the three active neutrinos is strongly
constrained by the primordial nucleosynthesis \cite{Okada,Bilenkii}.
The requirement $N_{\nu} < 4$ at neutrino decoupling
does not allow large mixing of a sterile neutrino with the active neutrinos,
and so   $\nu_{e}\leftrightarrow \nu_{s}$
and $\nu_{\mu}\leftrightarrow \nu_{\tau}$ oscillations are favored
for the solar and atmospheric neutrino deficit, respectively.
The implication of this conclusion on our gravitino-neutrino mixing is that
the SUSY breaking scale should be  $\sqrt{F} \leq 2\, \mbox{TeV}$.

In conclusion we have pointed out that the gravitino in the low energy SUSY
breaking models could mix with neutrinos through its spin half longitudinal
component and a lepton number violating bilinear term in the superpotential.
To be compatible with the neutrino experiments  the
SUSY breaking scale is required to be  in the range $ .2 \,\mbox{TeV}\leq 
\sqrt{F} \leq 20\, \mbox{TeV}$, and the lepton number violating term was
found to be small enough to evade the cosmological constraint.

\vspace{.5in}
\noindent
{\bf Acknowledgements:}
It is great pleasure to thank E.J. Chun for useful discussions and introducing
me his paper \cite{Chun}. This work was in part supported by the Korean
Science and Engineering Foundation (KOSEF).

\end{document}